\documentclass[final]{ws-rv9x6}

\usepackage{ws-rv-thm} 
\usepackage{hyperref}
\usepackage{xcolor}
\usepackage{listings}
\usepackage{xspace} % spaces in commands
\usepackage{multirow,booktabs,tabularx,float}
\usepackage{wrapfig}

\definecolor{dark-red}{rgb}{0.8, 0.0, 0.1803921568627451}
\definecolor{dark-blue}{rgb}{0.0, 0.0, 0.803921568627451}
\definecolor{dark-green}{rgb}{0.0, 0.39215686274509803, 0.0}
\definecolor{dark-orange}{rgb}{0.8, 0.4, 0.0}

\usepackage[square]{ws-rv-van}

\newcolumntype{R}{>{\raggedleft\arraybackslash}X}%
\newcolumntype{L}{>{\raggedright\arraybackslash}X}%

\newcommand{\toolfont}[1]{\texttt{#1}}

\newcommand{\be}{\begin{equation}}
\newcommand{\ee}{\end{equation}}

\newcommand{\diff}{\mathrm{d}}
\newcommand{\eg}{{e.\,g.}~}
\newcommand{\ie}{{i.\,e.}~}
\newcommand{\etc}{{etc.}~}

\newcommand{\thetaref}{\theta_{\text{ref}}}

\newlength{\hhatheight}

\DeclareMathOperator{\Pois}{Pois}

\DeclareMathOperator*{\argmax}{arg\,max}

\makeindex
\makeatletter\@addtoreset{chapter}{part}\makeatother%
\begin{document}

\setcounter{chapter}{6}
\chapter*{Simulation-based inference methods for particle physics}
\label{ch:gnn}

\author{Johann Brehmer and Kyle Cranmer}
\address{New York University, New York, NY, 10003}

\begin{abstract}
    Our predictions for particle physics processes are realized in a chain of complex simulators. They allow us to generate high-fidelty simulated data, but they are not well-suited for inference on the theory parameters with observed data. We explain why the likelihood function of high-dimensional LHC data cannot be explicitly evaluated, why this matters for data analysis, and reframe what the field has traditionally done to circumvent this problem. We then review new simulation-based inference methods that let us directly analyze high-dimensional data by combining machine learning techniques and information from the simulator. Initial studies indicate that these techniques have the potential to substantially improve the precision of LHC measurements. Finally, we discuss probabilistic programming, an emerging paradigm that lets us extend inference to the latent process of the simulator. 
\end{abstract}
%\markright{Customized Running Head for Odd Page} % default is Chapter Title.

\tableofcontents

\body

%============================================================
\section{Particle physics measurements as a simulation-based inference problem}
%============================================================

%------------------------------------------------------------
\subsection{A fundamental problem for LHC measurements}
%------------------------------------------------------------

Among the sciences, particle physics has the luxury of having a very well established theoretical basis. Quantum field theory provides a framework not only for the Standard Model, but also for theories of physics beyond the standard model (BSM). We almost take for granted the predictive power of our theories, but the way our field formulates searches for new new physics in terms of hypothesis tests and confidence intervals is critically tied to the fact that we have predictive models to test in the first place. 

Often we seem to equate the predictions of a theory with Feynman diagrams and the matrix element for a hard scattering process, which in turn can be used to predict a fully differential cross-section. Of course, that is not the full story, as one must include parton density functions and quarks and gluons give rise to a parton shower and subsequent hadronization process. Moreover, we observe electronic signatures tied to scintillation, ionization, \etc in our detectors, not the final-state particles directly. Therefore the predictive model for a theory must incorporate the response of the detector to the final state particles. 

While all of these points are well known to an experimental particle physicist, it has not been customary to describe the full simulation chain explicitly as a probabilistic model for the data. Why is that? In part that is because we have no explicit closed-form equation to write down nor do we have a function that we can provide to \toolfont{Minuit}~\cite{James:1975dr} that describes the probability distribution for the raw data in terms of parameters that appear in the Lagrangian for a given theory. Nevertheless, we can produce synthetic data using Monte Carlo simulation tools like \toolfont{MadGraph}~\cite{Alwall:2014hca}, \toolfont{Sherpa}~\cite{Gleisberg:2008ta}, \toolfont{Pythia}~\cite{Sjostrand:2014zea}, \toolfont{Herwig}~\cite{Corcella:2000bw}, and \toolfont{GEANT4}~\cite{Agostinelli:2002hh}.

%------------------------------------------------------------
\begin{table}
  \footnotesize
  \centering
  \tbl{Dictionary of symbols that appear in this review (derived from  Ref.~\cite{Brehmer:2018eca}).}{
      \begin{tabular}{p{.17\linewidth}@{\;}p{.5\linewidth}@{\;}p{.26\linewidth}}
        \toprule
        Symbol & Meaning & ML abstraction \\
        \midrule
        $\theta$ & Theory parameters & Parameters of interest \\
        $x$ & All observables & Features \\
        $v$ & 1-2 selected kinematic variables & Summary statistics \\
        $z_p$ & Parton-level four-momenta & Latent variables \\
        $z_s$ & Parton shower history & Latent variables \\
        $z_d$ & Detector interactions & Latent variables \\
        $z = (z_p, z_s, z_d)$ & Full simulation history of event & All latent variables \\
        \midrule
        $p_\mathrm{full}(\{x\}|\theta)$ & Full likelihood function, see Eq.~\eqref{eq:full_likelihood} & Implicit density \\
        $p(x | \theta)$ & Kinematic likelihood for single event & Implicit density \\[0cm]
        & (normalized fully differential xsec, Eq.~\eqref{eq:kin_likelihood}) &  \\
        $p_p(z_p | \theta)$ & Parton-level distribution & Tractable density \\
        $p_s(z_s| z_p)$ & Parton-shower effects & Implicit density \\
        $p_d(z_s| z_p)$ & Detector effects & Implicit density \\
        $p_x(x| z_d)$ & Detector readout & Implicit density \\
        $r(x | \theta)$ & Likelihood ratio function, see Eq.~\eqref{eq:likelihood_ratio} & \\
        $r(x,z|\theta)$ & Joint likelihood ratio, see Eq.~\eqref{eq:joint_likelihood_ratio} & Unbiased est.\ of $r(x|\theta)$ \\
        $t(x)$ & Score (locally optimal obs., Eq.~\eqref{eq:score}) \\
        $t(x,z|\theta)$ & Joint score, see Eq.~\eqref{eq:joint_score} & Unbiased est.\ of score \\
        \midrule
        $\hat{\theta}$ & Best fit for theory parameters & Estimator for $\theta$ \\
        $\hat{p}(x | \theta)$ & Parameterized estimator for likelihood \\
        $\hat{r}(x | \theta)$ & Parameterized estimator for likelihood ratio \\
        $\hat{s}(x | \theta)$ & Parameterized classifier decision function \\
        $\hat{t}(x)$ & Estimator for score \\
        $\hat{p}_{tf}(x | z_p)$ & Approximate shower and detector effects \\
        & (transfer function) \\
         \bottomrule
      \end{tabular}
  }
  \label{tbl:dictionary}
\end{table}
%------------------------------------------------------------

The colloquial term or jargon for both the simulation tools and the synthetic data they produce is \textit{Monte Carlo}. This term refers to methods that sample from probability distributions to compute an integral. Particle physics simulators use such methods to compute the cross section of a process by sampling a number of events following the probability distribution
\begin{equation}
    p(x, z_d, z_s, z_p | \theta) = p_\mathrm{x}(x | z_d) \, p_\mathrm{d}(z_d | z_s) \, p_\mathrm{s}(z_s | z_p) \, p(z_p | \theta) \,.
    \label{eq:joint_likelihood}
\end{equation}
Here the vector $z_p$ is the parton-level phase-space point of a simulated event (\ie the parton four-momenta, helicities, and charges); the vector $z_s$ summarizes the parton shower simulation, including the stable particles that emerge from it; $z_d$ are the interactions in the detector. These three vectors collectively define the ``Monte-Carlo truth record'' of a simulated event and are the \emph{latent variables} of the process: we cannot measure them, and in fact they are only well defined within a given simulator code. Finally, $x$ is the vector of observables. While a real-life observation consists of tens of millions of sensor read-outs, one can consider the  reconstruction of the event as part of the measurement process and take $x$ as a vector of four-momenta and other properties of the reconstructed particles. In Tbl.~\ref{tbl:dictionary} we provide a look-up table for these and other symbols that appear in this review and translate between particle physics and machine learning or statistics nomenclature.

There is an established chain of high-fidelity simulators that can sample events from the probability density in Eq.~\eqref{eq:joint_likelihood}. However, statistical inference---quantifying the degree to which parameter values $\theta$ are in agreement with an observed set of events $\mathcal{D} = \{x_i\}_{i=1}^n$---is surprisingly challenging. Why? The key quantity for both frequentist and Bayesian inference method is the likelihood function $p_\mathrm{full}(\mathcal{D}|\theta)$, the probability density of an observed set of events $\mathcal{D}$ as a function of the parameters $\theta$. The full likelihood function is given by

\begin{equation}
    p_\mathrm{full}(\mathcal{D}|\theta) = \Pois(n| \epsilon \, L \, \sigma(\theta)) \; \prod_i p(x_i | \theta) \,,
    \label{eq:full_likelihood}
\end{equation}
where $\Pois(n | \epsilon  \, L \, \sigma(\theta))$ is the Poisson probability density for $n$ observed events, efficiency and acceptance factors $\epsilon$, integrated luminosity $L$, total cross section $\sigma(\theta)$, and where
\begin{equation}
    p(x|\theta) = \int \! \diff z_d \int \! \diff z_s  \int \! \diff z_p \; p(x, z_d, z_s, z_p | \theta)
    \label{eq:kin_likelihood}
\end{equation}
is the probability density for an individual event to have data $x$. This  likelihood function involves integrals over the entire parton-level phase space, all possible shower histories, and all possible detector interactions compatible with the measurement $x$. The integral over this enormous space clearly cannot be computed in practice, so we cannot directly evaluate the likelihood of an observed event under different parameter values $\theta$. This means that we cannot directly find the maximum-likelihood estimators that best fit a given observation, construct confidence limits based on a likelihood ratio test statistic, or compute the Bayesian posterior $p(\theta | x)$!

The task of performing statistical inference when the data generating process does not have a tractable likelihood is known 
as \emph{simulation-based} or \emph{likelihood-free inference}. This case is not at all unique to particle physics. The formulation of this problem in a common, abstract language has led to statisticians, computer scientists, and domain scientists from various fields developing powerful methods for simulation-based inference together. While this review focuses on the particle physics case, the methods apply equally to a range of problems for instance in neuroscience, cosmology, or epidemiology.

%------------------------------------------------------------
\subsection{Solving the problem with summary statistics}
\label{sec:summary_stats}
%------------------------------------------------------------

If the intractability of the likelihood function is such a problem, how have high-energy physicists successfully analyzed particle collisions for decades? The reason that this problem is rarely acknowledged explicitly is that particle physicists have a track record of developing a good intuition about processes they study and finding powerful summary statistics for them. Summary statistics are individual variables that condense a high-dimensional observation. Typical examples are the reconstructed mass of a decaying unstable particle, decay angles between decay products, or other kinematic variables~\cite{Barr:2010zj, Jackson:2016mfb}. An ideal summary statistics vector $v$ captures  all of the relevant information in the observed event $x$ relevant to the parameter $\theta$, while being of much lower dimensionality. Given one or two summary statistics, we can easily compute the likelihood function $p(v|\theta)$ with histograms, kernel density estimation, or other density estimation techniques and then find the maximum-likelihood estimator in the parameter space and construct confidence limits based on the (profile) likelihood ratio test statistic~\cite{Diggle1984MonteCM, Cranmer:2000du}. This approach has been the workhorse of statistical analysis in collider physics for decades.

Note that most uses of machine learning in experimental particle physics take place within this approach. Experimental particle physicists have embraced the use of multivariate models (commonly boosted decision trees or fully connected neural networks) in the event selection. The statistical analysis of the events that pass this selection is then still based on histograms of kinematics-based summary statistics or the neural network output itself.

The reduction of data to summary statistics also enables Approximate Bayesian Computation (ABC)~\cite{rubin1984, beaumont2002approximate}, a simulation-based inference method that is gaining popularity in cosmology and is widely used in many scientific fields outside of physics. It directly targets Bayesian inference, using repeated runs of the simulator together with an accept-reject criterion to draw parameter samples that approximately follow the posterior.

Both the histogram method popular in particle physics and ABC suffer from the curse of dimensionality: the number of simulations required scales exponentially with the dimensionality of $x$ or $v$. This is why they only work with a low-dimensional statistic $v$ and cannot be effectively applied to high-dimensional data $x$. However, finding suitable summary statistics is a difficult and task-dependent problem and almost any choice of summary statistics discards some information. As a result, data analysis based on summary statistics typically leads to reduced sensitivity and statistical power.

%------------------------------------------------------------
\subsection{The frontier of simulation-based inference}
\label{sec:frontier}
%------------------------------------------------------------

In the next sections, we will describe modern simulation-based inference methods that allow us to analyze higher-dimensional data, improve the quality of inference, and improve the sample efficiency. Three developments are the key drivers behind these improvements~\cite{Cranmer:2019eaq}:
\begin{enumerate}
    \item The revolution in machine learning provides us with powerful surrogate models for the likelihood, likelihood ratio, or posterior function, or for optimal summary statistics. We can thus tap into the ability of modern machine learning methods to learn useful representations directly from high-dimensional data.
    \item Active learning methods iteratively use past results to steer the next simulations, leading to a better sample efficiency.
    \item Integrating inference capabilities with the simulation code and augmenting the training data with additional information that can be extracted from the simulator can substantially improve sample efficiency and quality of inference.
\end{enumerate}

Against the backdrop of these three broad trends, many different inference algorithms have been proposed in recent years, see Ref.~\cite{Cranmer:2019eaq} for an overview. Here we focus on a few methods that are particularly relevant for particle physics. In Sec.~\ref{sec:surrogates} we discuss techniques that aim to estimate the likelihood function or the likelihood ratio function, ranging from the Matrix-Element Method to machine learning--based methods to techniques that bring together matrix-element information and machine learning. Section~\ref{sec:optimal_observables} covers methods that aim to define powerful summary statistics, from parton-level Optimal Observables to neural network surrogates for the score function. We summarize and compare the main methods we discuss in Tbl.~\ref{tbl:comparison}. In the following sections we will briefly discuss diagnostic tools and systematic uncertainties as well as software implementations of these ideas. In Sec.~\ref{sec:prob_prog} we focus more on the latent process of the simulators and describe the paradigm of probabilistic programming. We discuss implementations of these methods in the HEP software stack in Sec.~\ref{sec:software}, before concluding with a summary in Sec.~\ref{sec:summary}.

%------------------------------------------------------------
\begin{table}
    \centering
    \scriptsize
    \tbl{Simulation-based inference methods for particle physics. We classify methods by the key quantity that is estimated in the different approaches, by whether they rely on a manual choice of summary statistics, are based on a transfer-function approximation (``TF''), whether their optimality depends on a local approximation (``local''), by whether they use any other functional approximations such as a histogram binning or a neural network (``NN''), whether they leverage matrix-element information (``$|\mathcal{M}|^2$''), and by the computational evaluation cost. Derived from a table in Ref.~\cite{Brehmer:2019bvj}.}{
        \begin{tabular*}{\textwidth}{@{\extracolsep{\fill}} l c c@{\,}c@{\,}c@{\, }c c c}
            \toprule
            \multirow{2}{*}{Method} & \multirow{2}{*}{Estimates} & \multicolumn{4}{c}{Approximations}
            & \multirow{2}{*}{$|\mathcal{M}|^2$} & \multirow{2}{*}{Comp.~cost} \\
            \cmidrule{3-6}
            & & summaries & TF & local & functional \\
            \midrule
            Histograms & $\hat{p}(v|\theta)$ & $\checkmark$ & & & binning & & curse of dim.  \\
            ABC & $\theta \sim p(\theta|v)$ & $\checkmark$ & & & $\epsilon$-kernel & & curse of dim. \\
            \midrule
            MEM & $\hat{p}(x|\theta)$ & & $\checkmark$ & & integral & $\checkmark$ & high (TF) \\
            NDE & $\hat{p}(x|\theta)$ & & & & NN & & amortized \\
            SCANDAL & $\hat{p}(x|\theta)$ & & & & NN & $\checkmark$ & amortized \\
            CARL & $\hat{r}(x|\theta)$ & & & & NN & & amortized \\
            RASCAL etc & $\hat{r}(x|\theta)$ & & & & NN & $\checkmark$ & amortized \\
            \midrule
            OO & $\hat{t}(x)$ & & $\checkmark$ & $\checkmark$ & integral & $\checkmark$ & high (TF) \\
            SALLY & $\hat{t}(x)$ & & & $\checkmark$ & NN & $\checkmark$ & amortized \\
            \bottomrule
        \end{tabular*}
    }
    \label{tbl:comparison}
\end{table}
%------------------------------------------------------------

%============================================================
\section{Inference with surrogates}
\label{sec:surrogates}
%============================================================

The first class of methods that we discuss tackles the problem head-on and constructs an estimator for the likelihood function $p(x|\theta)$ or the closely related likelihood ratio function
\begin{equation}
    r(x|\theta) = \frac {p(x|\theta)}{p_\mathrm{ref}(x)} \,,
    \label{eq:likelihood_ratio}
\end{equation}
where the denominator is some reference distribution, for instance using a reference value of the parameter points such as the Standard Model, a model average of multiple parameter points, or uniform phase space.

Once we have such an estimator, which we will denote $\hat{p}(x|\theta)$ or $\hat{r}(x|\theta)$, we can immediately use it in the established statistical pipeline: we can find the maximum-likelihood estimator for instance as
\begin{equation}
    \hat{\theta}_\mathrm{MLE} = \argmax_\theta \Pois(n| L \, \sigma(\theta)) \; \prod_i \hat{r}(x_i | \theta) 
\end{equation}
and similarly construct exclusion limits based on asymptotic properties of the (profile) likelihood ratio~\cite{Cranmer:2015nia}. Additionally, we can use the resulting likelihood ratio test statistic together with toy Monte Carlo to guarantee coverage, as discussed in Sec.~\ref{sec:systematics}.

%------------------------------------------------------------
\subsection{An approximation: the Matrix-Element Method}
\label{sec:mem}
%------------------------------------------------------------

The Matrix-Element Method (MEM)~\cite{Kondo:1988yd, Abazov:2004cs, Artoisenet:2008zz, Gao:2010qx, Alwall:2010cq, Bolognesi:2012mm, Avery:2012um, Andersen:2012kn, Campbell:2013hz,  Artoisenet:2013vfa, Gainer:2013iya, Schouten:2014yza, Martini:2015fsa, Gritsan:2016hjl, Martini:2017ydu, Kraus:2019qoq} approximates the likelihood in Eq.~\eqref{eq:kin_likelihood} by replacing the precise model of the effects of shower and detector with a simple, tractable transfer function $\hat{p}_{tf}(x | z_p)$. This simplifies the marginal distribution that would involve integrating over a large number of microscopic interactions to a convenient probability density such as a Gaussian. The MEM likelihood is given, schematically, by
\begin{equation}
  \hat{p}_{MEM}(x|\theta) = \int\!\diff z_p \; \hat{p}_{tf}(x | z_p) \, p(z_p | \theta)
  \sim \frac 1 {\sigma(\theta)} \;
  \int\!\diff z_p \; \hat{p}_{tf}(x | z_p) \, |\mathcal{M}(z_p | \theta)|^2 \,,
  \label{eq:mem}
\end{equation}
where $|\mathcal{M}(z|p | \theta)|^2$ is the squared matrix element evaluated at a phase-space point $z_p$ and parameters $\theta$ and for simplicity we have left out parton densities as well as phase-space and efficiency factors. Since the integrand is tractable and the integral is over a much lower-dimensional space than the on in Eq.~\eqref{eq:kin_likelihood}, it is feasible---though expensive---to compute this approximate likelihood function.\footnote{This approach has also been extended to include an explicit calculation of leading parton-shower effects~\cite{Soper:2011cr, Soper:2012pb, Soper:2014rya, Englert:2015dlp}.}

In some processes, particularly those involving only leptons and photons, the MEM can give a reliable estimate of the true likelihood.  However, jets are less well modeled by transfer functions and additional jet radiation is difficult to describe in this approach. Finally, the MEM still requires a computationally expensive numerical integration  \emph{for every event that is evaluated}, which can be prohibitive.

%------------------------------------------------------------
\subsection{Learning the likelihood}
\label{sec:likelihood}
%------------------------------------------------------------

Rather than computing the integral in the likelihood for every event to be evaluated, we can fit a surrogate model to data from the simulator and then use that for inference. Such a surrogate model needs to be flexible enough to accurately represent a complicated and multimodal probability distribution, we have to fit it to limited training data, and its likelihood function needs to be computed efficiently. Kernel density estimation has been used in this context~\cite{Holmstrom:1995bt}, but it was limited to roughly five-dimensional data. Recently, several machine learning models have been developed for this task, which are effective for estimating distributions of high-dimensional data. In particular, neural density estimators such as normalizing flows~\cite{2015arXiv150505770J, 2019arXiv191202762P} are flexible probabilistic models with a tractable likelihood function.

This lets us solve the problem of simulation-based inference in three phases~\cite{Cranmer:2016lzt}:
\begin{enumerate}
    \item We run the usual simulator chain a number of times with different input parameters $\theta$ and saving $\theta$ together with the simulated events $x \sim p(x|\theta)$.
    \item Next, a neural density estimator is trained to learn the conditional probability density $p(x|\theta)$. We use a single model for the whole parameter space (as opposed to individual models for a number of points along a grid in the parameter space), the parameter point $\theta$ to be evaluated is an additional input to the model. Such a \emph{parameterized} model~\cite{Cranmer:2015bka, Baldi:2016fzo} can leverage the smooth dependence on the parameter space, the probability density at each parameter point can ``borrow'' information from nearby points).
    \item After training, we can evaluate this model for arbitrary observations $x$ and parameter points $\theta$ and efficiently get an estimator for the likelihood function $\hat{p}(x|\theta)$. We can then use this to define best-fit points and exclusion limits with the usual statistical tools.
\end{enumerate}

Two aspects of this approach are noteworthy. First, we can use any state-of-the-art simulator in this approach, including shower and detector effects. Unlike in the MEM, there is no need for any approximations on the underlying physics. Second, the approach is \emph{amortized}: after an upfront simulation and training phase, we can evaluate the approximate likelihood function very efficiently for a large number of events and parameter values.  

Neural density estimators like normalizing flows have other useful properties. They are generative models, \ie one can sample from the probability distributions they have learned. This is not only a convenient cross check, but can also be used for event generation, to unfold reconstruction-level variables to the parton level~\cite{Bellagente:2020piv}, and for anomaly detection~\cite{Nachman:2020lpy}.

%------------------------------------------------------------
\subsection{Learning the likelihood ratio}
\label{sec:likelihood_ratio}
%------------------------------------------------------------

Training a surrogate for the likelihood function actually solves a harder problem then necessary for inference. To find the maximum-likelihood parameter point and to construct exclusion limits we do not actually need to know the likelihood function itself---the likelihood \emph{ratio} $r(x|\theta)$ defined in Eq.~\eqref{eq:likelihood_ratio} is in fact just as useful! As it turns out, training a neural network to learn the likelihood ratio  is often easier than learning the likelihood function.

The key idea is known as the \emph{likelihood ratio trick}: a binary classifier trained to discriminate samples $x \sim p(x|\theta)$ from samples $x \sim p_\mathrm{ref}(x)$ will eventually\footnote{In the limit of a sufficiently expressive model, infinite training data, and efficient minimization of the loss function.} converge to the output $\hat{s}(x|\theta) \to p_\mathrm{ref}(x) / [p(x|\theta) + p_\mathrm{ref}(x)]$, which is a monotonic function of the likelihood ratio $r(x|\theta)$. In other words, we can transform the output of a classifier $\hat{s}(x|\theta)$ into an estimator for the likelihood ratio function as
\begin{equation}
    \hat{r}(x|\theta) = \frac {1 - \hat{s}(x|\theta)} {\hat{s}(x|\theta)} \,.
    \label{eq:likelihood_ratio_trick}
\end{equation}

We can use this in an inference algorithm similar to the one discussed in the previous section~\cite{Neal:2007zz, 2012arXiv1212.1479F, Cranmer:2015bka, cranmer_2016, Louppe:2016aov, 2016arXiv161003483M, 2016arXiv161110242D, gutmann2017likelihood, 2018arXiv181009899D, Hermans:2019ioj, Andreassen:2019nnm}:
\begin{enumerate}
    \item We again start by running the simulator chain, generating one set of events from a reference distribution $p_\mathrm{ref}(x)$ (\eg the SM) and a second set of events from various parameter points $\theta$.
    \item Next, a neural classifier is trained to discriminate between these two sets, using the binary cross-entropy as a loss function. Like before, the classifier is parameterized: the parameters $\theta$ are used as explicit inputs into the classifier.
    \item After training, we can transform the output of the classifier into an estimator for the likelihood ratio function with Eq.~\eqref{eq:likelihood_ratio_trick} and an optional calibration procedure. This surrogate model can then be used to find the best-fit point and exclusion contours using established statistical tools.
\end{enumerate}

This method is known as CARL~\cite{Cranmer:2015bka, Louppe:2016aov, leonora_vesterbacka_2020_4062049}. It again supports arbitrary simulators without requiring approximations on the underlying physics and is amortized, allowing for an efficient evaluation after an upfront simulation and training cost. Compared to learning the likelihood function with a neural density estimator, the CARL approach can be more sample efficient (saving computation time).

While a surrogate model for the likelihood ratio does not allow us to generate samples, it can be used for reweighting. For the simulation-based inference problem, this can be useful as a diagnostic tool. In other contexts, this ability can be used to reweight events~\cite{cranmer_2016, Andreassen:2019nnm, leonora_vesterbacka_2020_4062049}, tune shower and detector-simulation parameters to data~\cite{Andreassen:2019nnm}, for unfolding~\cite{Andreassen:2019cjw}, and for anomaly detection~\cite{Andreassen:2020nkr}.

%------------------------------------------------------------
\subsection{Integration and augmentation}
\label{sec:mining_gold}
%------------------------------------------------------------

Both inference techniques described above---training a neural density estimator to learn the likelihood function and training a classifier to learn the likelihood ratio---treat the simulator chain as a black box that takes parameters $\theta$ as input and outputs samples $x \sim p(x|\theta)$. In reality, though, we know more about the particle physics processes. They consist of the separate pieces of parton-level generator, parton shower, and detector simulation, as given by Eq.~\eqref{eq:kin_likelihood}; typically only the parton-level step explicitly depends on the theory parameters of interest $\theta$.

We can leverage this understanding of the simulated process to extract more information from the simulator and use it to augment the training data for the likelihood or likelihood ratio model.\footnote{The extraction of the joint likelihood ratio and score is in fact more general and can be realized for many simulators~\cite{Brehmer:2018hga, Brehmer:2019jyt}. However, the particular structure of particle physics processes makes it easy to compute these quantities, which in this case are closely linked to the squared matrix element.} In particular, we can access the latent variables (or Monte-Carlo truth record) $z = (z_p, z_s, z_d)$, while tools like \toolfont{MadGraph} let us compute matrix elements for arbitrary theory parameters. For each simulated event we can thus compute two useful quantities: the \emph{joint likelihood ratio}~\cite{Brehmer:2018hga, Brehmer:2018kdj, Brehmer:2018eca}
\begin{align}
  r(x,z | \theta)
  &\equiv \frac {p(x, z | \theta)} {p_\mathrm{ref}(x, z)} \notag \\
  &= \frac {p(x|z_d) \, p(z_d|z_s) \, p(z_s|z_p) \, p(z_p|\theta)} {p(x|z_d) \, p(z_d|z_s) \, p(z_s|z_p) \, p_\mathrm{ref}(z_p)} \notag \\
  &= \frac {|\mathcal{M}|^2(z_p | \theta)} {|\mathcal{M}|^2_\mathrm{ref}(z_p)} \, \frac {\sigma_\mathrm{ref}} {\sigma(\theta)}
  \label{eq:joint_likelihood_ratio}
\end{align}
and the joint score
\begin{align}
  t(x, z | \theta)
  &\equiv \nabla_\theta \log p(x, z | \theta) \notag \\
  &= \frac {p_x(x|z_d) \, p_d(z_d|z_s) \, p_s(z_s|z_p) \, \nabla_\theta p_p(z_p|\theta)}
  {p_x(x|z_d) \, p_d(z_d|z_s) \, p_s(z_s|z_p) \, p_p(z_p|\theta)} \notag \\
  &= \frac {\nabla_{\theta} |\mathcal{M}|^2(z_p | \theta)} {|\mathcal{M}|^2(z_p | \theta)}
  - \frac {\nabla_{\theta} \sigma(\theta)} {\sigma(\theta)} \,.
  \label{eq:joint_score}
\end{align}
Here $|\mathcal{M}|^2(z_p|\theta)$ and $|\mathcal{M}|^2_\mathrm{ref}(z_p)$ are the squared matrix elements for parton-level phase space points $z_p$ for theory parameters $\theta$ and under the reference distribution, respectively, while $\sigma(\theta)$ and $\sigma_\mathrm{ref}$ are the total cross sections. The joint likelihood ratio and joint score quantify how the probability of one simulated event---fixing all of the latent variables in the simulation chain---changes if we change the theory parameters $\theta$.

\begin{figure}[t]
    \centering
    \includegraphics[width=\textwidth]{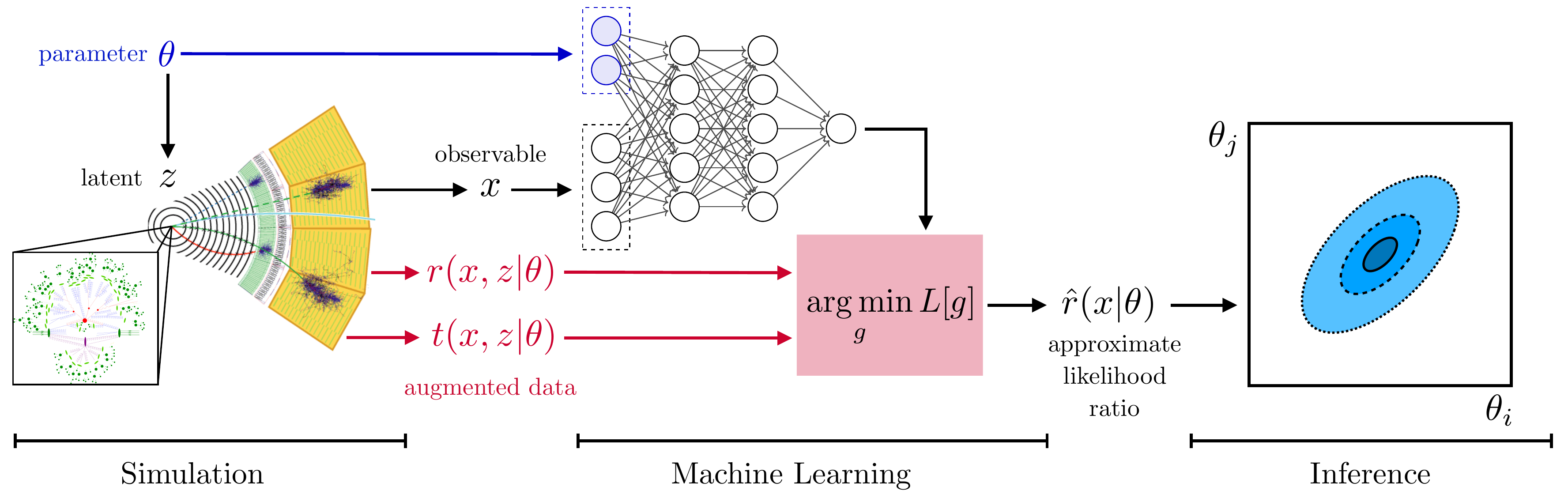}
    \caption{Illustration of a complete inference method using the RASCAL or ALICES methods to train a surrogate for the likelihood ratio function. Figure taken from Ref.~\cite{Brehmer:2018kdj}.}
    \label{fig:rascal_explainer}
\end{figure}

How are these two quantities useful, especially given that they depend on latent variables $z$ that are only meaningful for simulated events, but not for real measurements? It turns out that the joint likelihood ratio $r(x,z|\theta)$ is an unbiased estimator of the likelihood ratio $r(x|\theta)$ and the joint score provides unbiased gradient information. This means that we can augment our training data with these numbers and use them as labels in a supervised training setup. This idea is realized in a few different algorithms, which mostly differ by the loss functions that they use: the SCANDAL method improves the training of neural surrogates for the likelihood function~\cite{Brehmer:2018hga}, while the RASCAL~\cite{Brehmer:2018hga, Brehmer:2018kdj, Brehmer:2018eca} and ALICES~\cite{Stoye:2018ovl} techniques train neural surrogates of the likelihood ratio function more efficiently. After training the surrogates for the likelihood or likelihood ratio, we are again left with a neural network that can be evaluated for arbitrary events and parameter points and allows for amortized inference as before. The full workflow is schematically shown in Fig.~\ref{fig:rascal_explainer}.

\begin{figure}[t]
    \centering%
    \includegraphics[width=\textwidth]{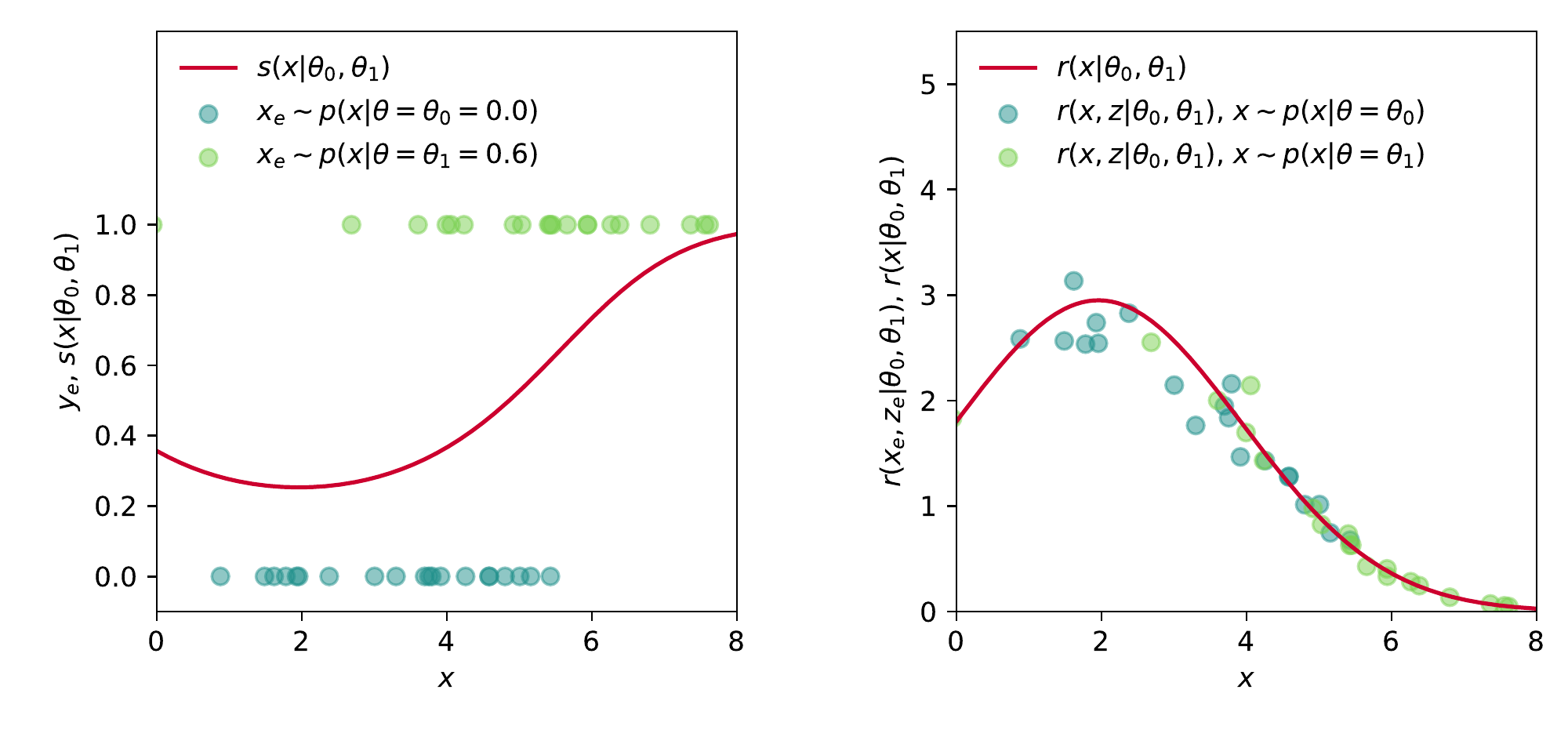}\\
    \includegraphics[width=0.5\textwidth,trim={10.4cm 0 0 0},clip]{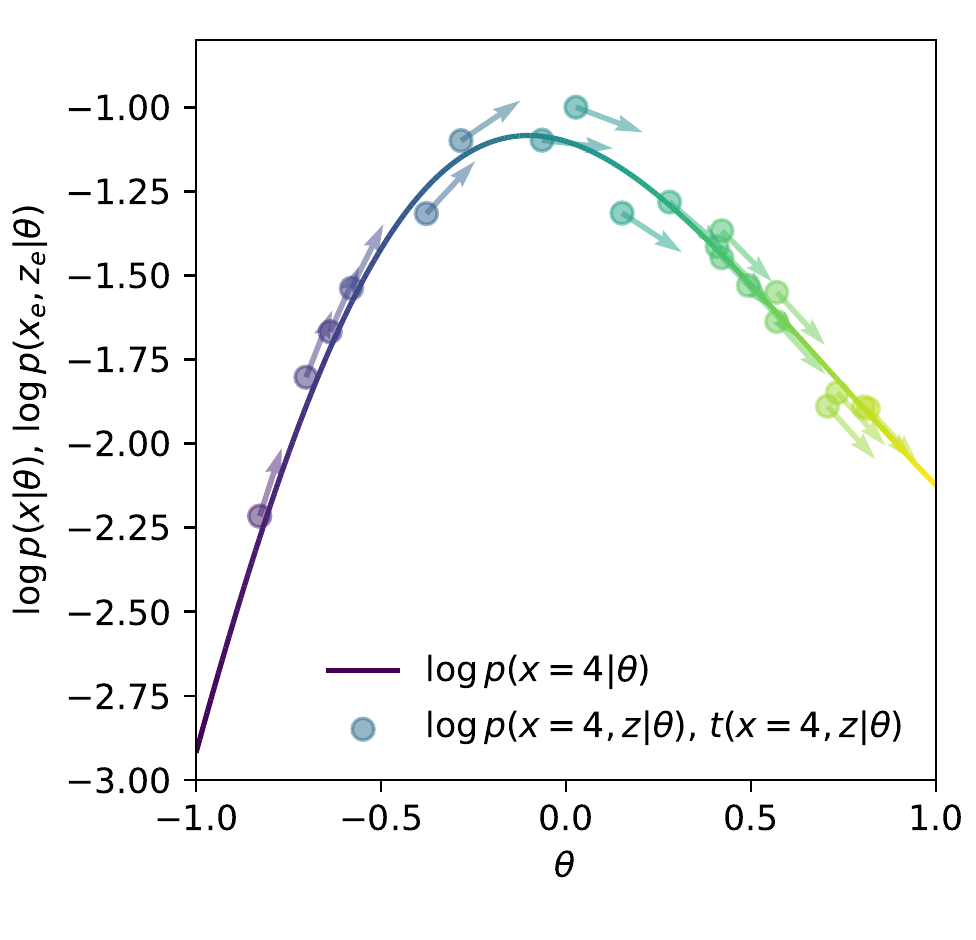}%
    \includegraphics[width=0.5\textwidth,trim={12.5cm 0 0.3cm 0},clip]{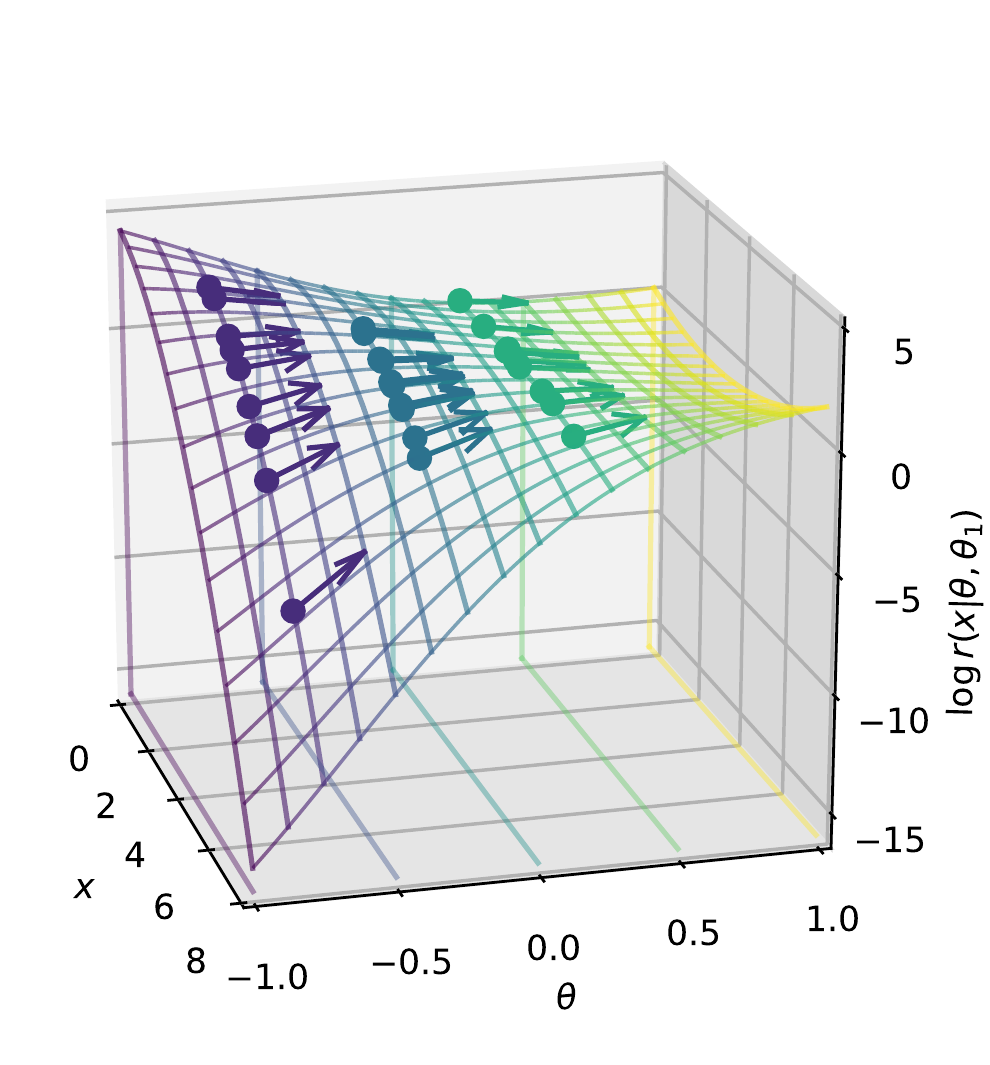}%
    \caption{Illustration of different approaches to train surrogates for the likelihood ratio function and the role of joint likelihood ratio and joint score. Figures taken from Ref.~\cite{Brehmer:2018eca}. \textbf{Top left}: in the likelihood ratio trick and the CARL inference method, a classifier decision function (red) has to be learned from binary labels that are zero or one (green dots). \textbf{Top right}: the joint likelihood ratio provides noisy, but unbiased labels (green) for the likelihood ratio function to be learned (red). \textbf{Bottom left}: the joint score adds noisy, but unbiased gradient information (arrows). \textbf{Bottom right}: the RASCAL and ALICES methods combine three orthogonal pieces of information (dots with arrows), allowing a neural network to learn the likelihood ratio function (surface) more efficiently.}
    \label{fig:illustration_latent}
\end{figure}

Extracting the joint likelihood ratio and joint score during the simulation stage and augmenting the training data with it adds multiple orthogonal pieces of information to the training, as we illustrate in Fig.~\ref{fig:illustration_latent}. In practice, this substantially reduces the number of simulated events that are necessary for a good performance---in some cases by multiple orders of magnitude~\cite{Brehmer:2018kdj, Brehmer:2018eca}!

Some particle physics measurements have even more additional structure, for instance when we are trying to constrain the Wilson coefficients of an effective field theory and the squared matrix element is a polynomial of these parameters. Incorporating this additional structure in the inference workflow can improve the efficiency even further~\cite{Brehmer:2018eca, ATLAS:morphing, Chen:2020mev}.

%------------------------------------------------------------
\subsection{Active learning}
\label{sec:active_learning}
%------------------------------------------------------------

Active learning here describes a sequential workflow that alternates simulation and inference stages. The theory parameters for which more events are generated are chosen such that they are expected to be most useful based on the observed data and the results of past iterations. Different algorithms have been proposed, some of which are based on neural surrogates for the likelihood~\cite{2018arXiv180507226P, 2018arXiv180509294L}, while others target the likelihood ratio~\cite{Hermans:2019ioj, 2020arXiv200203712D}. While active learning is often phrased in a Bayesian framework, these methods can be applied equally well to frequentist inference~\cite{ranjan2008sequential, Bect2012, lukas_heinrich_2018_1634428}.

Active learning maximizes sample efficiency for a particular observed data set. This is somewhat at odds with the goal of amortization, which aims to train a surrogate model that works well for multiple different data sets. While active learning can be very powerful in cases with few observed data points, it is less crucial in particle physics use cases with a large number of expected or observed events.

%============================================================
\section{Inference with sufficient summary statistics}
\label{sec:optimal_observables}
%============================================================

The methods discussed in the previous section tackle simulation-based inference by learning the likelihood or likelihood ratio function in the high-dimensional data space. While these methods are powerful, they require an analysis workflow that is substantially different from the high-energy physics standards. This makes modifications to the usual software pipeline, careful cross checks, and some changes to the way that systematic uncertainties are handled necessary (more on this later).

A more incremental change to the current analysis workflow is to construct powerful summary statistics in a systematic way. After the high-dimensional data $x$ for an event is compressed to one or a few of these summary statistics, it can be analyzed in the usual, histogram-centric way described in Sec.~\ref{sec:summary_stats}. The analysis workflow remains largely unchanged, except that instead of kinematic variables (like the transverse momentum of a jet) more complicated variables (like the output of a neural network) are analyzed. No essential modifications to the software pipeline or the treatment of systematic uncertainties are necessary in this approach.

So how do we find these optimal summary statistics? There are two broad strategies. The first is to try to learn a summary statistic as an intermediate step in the end-to-end analysis of the data where the objective function is, for instance, an expected significance or expected limit as in INFERNO~\cite{deCastro:2018mgh} or neos~\cite{lukas_heinrich_2020_3697981}. This is connected to the recent discussions around differentiable programming. Optimizing an experiment-level objective is computationally expensive, and not actually necessary since the data are independent and the likelihood factorizes as in Eq.~\ref{eq:full_likelihood}.%
\footnote{If we knew the full likelihood $p( \mathcal{D} |\theta, \nu)$ in Eq.~\eqref{eq:full_likelihood}, where $\theta$ are parameters of interest and $\nu$ are nuisance parameters, the final test statistic we would target would be the profile likelihood ratio $\lambda(\theta) = p(\mathcal{D} | \theta, \hat{\hat{\nu}} ) / p(\mathcal{D} | \hat{\theta}, \hat{\nu})$, where $\hat{\theta}$ and $\hat{\nu}$ are the maximum likelihood estimator (MLE) and $\hat{\hat{\nu}}$ is the conditional maximum likelihood estimator (CMLE)~\cite{Cowan:2010js}. The numerator and denominator of the likelihood of the likelihood ratio factorize across experiments, but the values for the MLE and CMLE couple all of the events in the dataset $\mathcal{D}$. However, this coupling of events through the MLE and CMLE can be postponed and based on a learned surrogate for the per-event likelihood or likelihood ratio function as discussed in the previous section.}

Alternatively, we can look for sufficient statistics that allow us to approximate the per-event likelihood, and there are many advantages to casting the learning problem in terms of individual events. While our exposition will focus on the parameters of interest, one can consider $\theta$ to also include nuisance parameters, and profiling the nuisance parameters would then happen down-stream in the statistical inference pipeline and after the amortized learning stage described below. 

The key to learning optimal observables is to consider a local approximation of the likelihood function in the parameter space. In other words, assume that we are studying parameters $\theta$ that are close to some chosen reference parameter point $\thetaref$ (imagine this, for instance, to be the Standard Model). Then one can show~\cite{Alsing:2017var, Brehmer:2018kdj, Alsing:2018eau} that the most powerful observable for measuring $\theta$ is the \emph{score}
\begin{equation}
    t(x) = \nabla_\theta \log p(x|\theta) \Bigr|_{\thetaref} \,.
    \label{eq:score}
\end{equation}
This gradient vector contains one component per parameter of interest. In the neighborhood of $\thetaref$, the score components are the sufficient statistics: analyzing just $t(x)$ will yield just as much information about $\theta$ than analyzing the high-dimensional data $x$. By using the score as summary statistics, we are therefore not throwing away any information, at least as long we focus on parameters close to $\thetaref$. Further away from the reference point, the score components may no longer be sufficient and a histogram-based analysis will no longer be optimal.

Unfortunately, like the likelihood function itself, the score is in general intractable. In the following we will present two methods that allow us to estimate it.

%------------------------------------------------------------
\subsection{An approximation: parton-level Optimal Observables}
\label{sec:classical_oo}
%------------------------------------------------------------

Remember that the Matrix-Element Method approximated the likelihood function by summarizing the effect of shower and detector with a transfer function. Parton-level Optimal Observables (OO)~\cite{Atwood:1991ka, Davier:1992nw, Diehl:1993br} use the same approximation to compute the score:
\begin{align}
    \hat{t}_{OO}(x)
    &= \nabla_\theta \log \left( \int\!\diff z_p \; \hat{p}_{tf}(x|z_p) \, p(z_p | \theta) \right) \Biggr |_{\theta = \thetaref} \notag \\
    &= \dfrac {\displaystyle \int\!\diff z_p \; \hat{p}_{tf}(x|z_p) \, \nabla_\theta p(z_p | \thetaref)}
      {\displaystyle \int\!\diff z_p \; \hat{p}_{tf}(x|z_p) \, p(z_p | \thetaref)} \,.
    \label{eq:parton_level_oo}
\end{align}
In practice, this method is usually applied to processes with easily identifiable final-state particles like leptons and photons. In that case, the reconstructed particle properties are simply identified with the parton-level four-momenta, $\hat{p}(x|z_p) = \prod_i \delta^{4}(x_i - z_{p\,i})$.

While this approach elegantly uses our knowledge of matrix elements, it requires substantial approximations to the underlying process, and taking into account shower or detector effects in the observable detection leads to a large computational cost for each analyzed event.

%------------------------------------------------------------
\subsection{Learning the score}
\label{sec:sally}
%------------------------------------------------------------

The SALLY method~\cite{Brehmer:2018hga, Brehmer:2018kdj, Brehmer:2018eca} trains a neural network to learn the (intractable) score function $t(x)$ including the full detector simulation. As in the methods discussed in Sec.~\ref{sec:mining_gold}, the first step is running the simulator chain a number of times, now using the reference parameter point $\thetaref$ as input. In addition to the observation $x$, the joint score $t(x,z|\thetaref)$ defined in Eq.~\eqref{eq:joint_score} is computed and stored for every simulated event. In a next step, a machine learning model like a neural network $\hat{t}(x)$ is trained to minimize the mean squared error $|\hat{t}(x) - t(x,z)|^2$. It can be shown that the neural network will ultimately converge to the score function given in Eq.~\eqref{eq:score}. After training, the neural network thus defines the locally most powerful observables for the measurement of $\theta$ and can be used in a standard analysis pipeline.

In addition to defining locally optimal observables, neural score estimators can also be used to compute the Fisher information, a versatile tool for sensitivity forecasting, cut optimization, and feature selection~\cite{Brehmer:2016nyr, Brehmer:2017lrt, Brehmer:2019xox}.

%============================================================
\section{Diagnostics, calibration, and systematic uncertainties}
\label{sec:systematics}
%============================================================

The analysis methods described in the previous sections contain some parts, in particular neural networks, that are not always easy to interpret and can be harder to debug than a standard analysis based on histograms of traditional observables. It is important to make sure that we can trust the results and quantify any systematic uncertainties. This is very similar to basing the downstream statistical analysis on histograms of neural network outputs. 

Mainly we have to correctly diagnose model misspecification. Inference is always performed within the context of a statistical model, but if that model is not correct for a task at hand, the inference results will be meaningless or misleading. In the simulation-based inference methods we discuss, two types of models appear, both of which are prone to misspecification: the simulator itself and machine learning surrogates. This is similar to the distinction between the full simulation and the use of an analytic function (surrogate) to model a smooth $m_{\gamma\gamma}$ spectrum in a $H\to \gamma\gamma$ analysis. 

Misspecification of the simulator occurs when \toolfont{MadGraph}, \toolfont{Pythia}, \toolfont{Geant4} \etc do not model the physics of LHC collisions accurately enough. This problem also plagues classical histogram-based analyses, but may be easier to diagnose and calibrate when only a single variable is studied than in the multivariate analysis methods described here~\cite{Nachman:2019yfl}. It is usually addressed by varying the parameters of the simulator, which introduces nuisance parameters $\alpha$ with unknown true values, and profiling over them in the statistical analysis. We can also use ideas from domain adaptation and algorithmic fairness to make the neural network less sensitive to variations in the nuisance parameters~\cite{Louppe:2016ylz, deCastro:2018mgh, Alsing:2019dvb}. If possible, however, it is conceptually cleaner to explicitly include the effect of nuisance parameters in the likelihood model $\hat{p}(x|\theta, \alpha)$ or $\hat{r}(x|\theta, \alpha)$ and to use well-defined and established statistical procedures like profiling to take them into account in the downstream statistical analysis.

Misspecification of the surrogate model occurs when the neural network  does not approximate the true likelihood or likelihood ratio perfectly. This is analogous to a falling exponential for the $m_{\gamma\gamma}$ spectrum not fitting the simulated data perfectly.  Typical reasons are the limited number of training samples, insufficient network capacity, or an inefficient minimization of the loss function. A common issue is that the classifier $\hat{s}(x|\theta)$ will be roughly one-to-one with the true likelihood ratio, but not exactly. This can be fixed with the calibration procedure used in CARL and described in Ref.~\cite{Cranmer:2015bka}. One can protect against more severe  deficiencies by calibrating the inference results with toy simulations from the simulator: for every parameter point, we can run the simulator to construct the distribution of the likelihood or likelihood ratio. Ultimately this leads to confidence sets with a coverage guarantee (assuming the simulator is accurate) as in the Neyman Construction, \ie that will never overly optimistic~\citep{Cranmer:2015bka, Brehmer:2018eca}. This toy Monte Carlo approach can require a large number of simulations, especially for high-dimensional parameter spaces. For an in-depth discussion of calibration and the Neyman construction, see Ref.~\cite{Brehmer:2018eca}.

There are other, less computationally expensive tools to diagnose misspecification of the surrogate model. These include off-the-shelf uncertainty quantification methods for neural networks such as ensemble methods and Bayesian neural networks. In addition, one can train classifiers to distinguish data from the surrogate model and the true simulator~\citep{Cranmer:2015bka}, check certain expectation values of estimators of the likelihood, likelihood ratio, or score against a known true value~\citep{Brehmer:2018eca}, vary unphysical reference distributions that should leave the inference result invariant~\citep{Cranmer:2015bka}, and compare the distribution of network outputs against known asymptotic properties~\citep{Wilks:1938dza, Wald, Cowan:2010js}. Passing these closure tests does not guarantee that a model is correct, but failing them is an indication of an issue.

%============================================================
\section{Probabilistic programming}
\label{sec:prob_prog}
%============================================================

\setlength{\columnsep}{10pt}%
\begin{wrapfigure}{r}{.48\textwidth}
\vskip-16pt
\begin{lstlisting}[language=Python,basicstyle=\small]
def stochastic_function():
    z1 = rand()
    if z1 < 0.5:
        z2t = rand()
        x = z1 + z2t
    else:
        z2f = rand()
        z3f = rand()
        x = z1 + z2f + z3f 
    return x
\end{lstlisting}
\vskip-2pt
\caption{Illustrative example of a stochastic function.}
\label{algo:stochastic_function}
\vskip-6pt
\end{wrapfigure}
We will now switch gears and review probabilistic programming, a set of methods that are related to, but different from the simulation-based inference techniques discussed in the previous sections. Computer programs that involve random numbers and do not have deterministic input-output relationships can be thought of as specifying a probability distribution $p(\textrm{output} | \textrm{input})$. It is natural to think of simulators in this way, where the parameters of the simulator are identified with $\theta$ and the output of the simulator is identified with $x$. Furthermore, the values of the random variables and the other intermediate quantities inside the computer program can be thought of as latent variables $z$. The structure of the space of latent variables can also be complex. Consider the simple example in Fig.~\ref{algo:stochastic_function}, where the list of latent variables is either \texttt{(z1, z2t)} or \texttt{(z1, z2f, z3f)} and depends on the control flow of the program.

It can be useful to think of the latent space of such a program as the space of its stack traces along with the values of all the variables.  Take a moment to think about the complexity of the typical simulation chain going from matrix elements to parton  shower and hadronization through the detector simulation. These programs have enormous, highly structured latent spaces. The probability that the program returns $x$ corresponds to integrating over all the possible executions of the program that could return $x$; as we argued in the introduction of this review, this is intractable for moderately complicated programs.

We saw in Sec.~\ref{sec:surrogates} how we can use machine learning surrogates to approximate the likelihood $p(x|\theta)$ or likelihood ratio $r(x|\theta)$, where the dependence on $z$ has been marginalized or integrated out. One of the advantages of those approaches is that the surrogate models don't attempt to capture the complexity of the latent state or the joint distribution $p(x,z|\theta)$. But what if we also want to infer something about the latent variables that describe what is going on inside the simulator?

In HEP it is common to inspect the Monte Carlo truth record (i.e. $z$) for some set of events that satisfy some cuts to gain insight into why something happens. For instance, we might want to know what happened inside the simulation of $pp\to j j$ events that led to very large missing transverse energy, or why a jet faked a muon. To study this, we often filter a large set of events (simulated with a particular parameter setting $\theta$), filter those events that satisfy the cuts, and then look at histograms of some particular Monte Carlo truth quantities $f(z)$ (for example, to inspect if there was a semi-leptonic $b$-decay or punch-through in the calorimeter). That familiar procedure is approximating the posterior distribution of $f(z)$ given that the event generated with parameter $\theta$ passes the cuts, which we can write symbolically as $p(f| \texttt{cuts}(x)=\texttt{True}, \theta)$. Similarly, the unfiltered sample can be thought of as samples from the prior $p(f| \theta)$. 

The problem with the traditional approach is that the filter efficiency can be very low, and very few of the prior samples may survive to estimate the posterior. This is similar to the inefficiency found in Approximate Bayesian Computation, which asks for the simulator to generate an a simulated $x$ close to the observed $x_\mathrm{obs}$. This motivates an additional language construct that allows for conditioning on random variables, which characterizes probabilistic programming. Probabilistic programming languages (PPLs) extend general-purpose programming languages with constructs to do sampling and conditioning of random variables~\cite{Gordon2014ProbabilisticP, wood-aistats-2014}%
\footnote{Often it is assumed that the quantity being conditioned on is directly sampled from a distribution with a known likelihood (conditioned on the latent state of the simulator at that point in the execution). Sometimes this is reasonable, but sometimes this assumption is violated and we want to condition on some more complicated function of the random variables with an intractable density. In that setting, one typically needs to introduce some tolerance or kernel. In this way, probabilistic programming can be seen as a more sophisticated and computationally efficient way of implementing  Approximate Bayesian Computation.}.

The additional language constructs express the concept of sampling and conditioning, but they do not necessarily specify how that is implemented. It is best to decouple the model specification (the probabilitic program or simulator code) from the inference algorithm---much as we use a tool like \texttt{HistFactory}~\cite{Cranmer:2012sba} to create a statistical model and then use \texttt{RooStats}~\cite{Moneta:2010pm} to provide generic statistical inference algorithms. Various inference engines have been developed implementing different inference strategies such as Importance Sampling~\cite{2019arXiv190703382G} and specializations of Metropolis-Hastings~\cite{pmlr-v15-wingate11a} that are compatible with the complex latent space structure associated to stack traces. In general, the inference algorithms can be thought of as hijacking the random sampling inside of the simulator code to guide the simulator towards a certain output. 

\begin{figure}[t]
    \includegraphics[width=\textwidth]{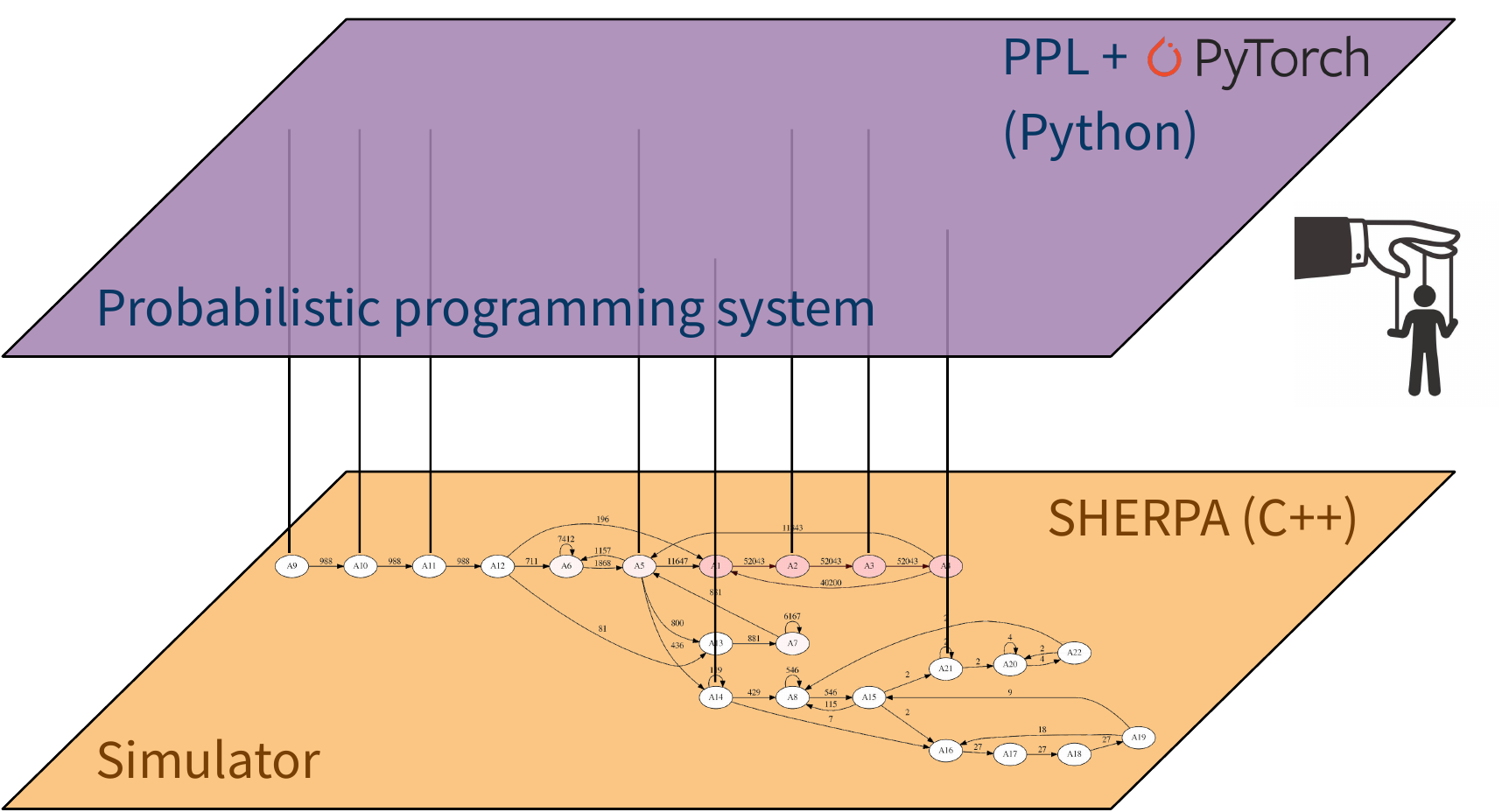}
    \caption{An illustration of a Python-based probabilistic programming system's inference engine controlling the \texttt{SHERPA} event generator through the \texttt{PPX} protocol. Figure taken from Ref.~\cite{baydin2019efficient}.}
    \label{fig:puppetmaster}
\end{figure}

Early research in probabilistic programming required coding the simulator in special-purpose languages, which is not an attractive option for HEP as we have decades of work invested in our simulation code bases. Recently, however, the Etalumis project developed \texttt{PPX}, a cross-platform probabilistic execution protocol that allows an inference engine to control a simulator in a language-agnostic way~\cite{2019arXiv190703382G,baydin2019efficient}. The Etalumis team integrated \texttt{PPX} into the \texttt{SHERPA} simulator and a simplified calorimeter simulation to demonstrate probabilistic programming with a real-world simulator  (see Fig.~\ref{fig:puppetmaster}). 

The bulk of the probabilistic programming literature is phrased in terms of Bayesian statistics. The posterior distribution $p(z|x, \theta)$ is of no conceptual problem for an ardent, frequentist particle physicists, because while $z$ may be latent, it is a random variable and the joint distribution $p(x,z | \theta)$ is perfectly well defined. However, if one wanted to use probabilistic programming to infer the parameters of the simulation $\theta$, then one would need to include a prior $p(\theta)$ and sample from that distribution at the beginning of the program. The result would be a probabilistic program for the joint model $p(x,z,\theta) = p(x,z|\theta) p(\theta)$, and one would then condition on $x$ to obtain samples from the posterior $p(\theta, z | x)$ or the the marginal  $p(\theta | x)$.

%============================================================
\section{Software and computing}
\label{sec:software}
%============================================================

The methods described in this review are closely connected to the software and computing challenges of high-energy physics, particularly when we think about the high-luminosity LHC. 

Initial results from phenomenological studies indicate that these new machine-learning based approaches provide substantial improvements in sensitivity to traditional approaches, but generating the training data is computationally expensive. However, with some additional work, the augmented data described in Sec.~\ref{sec:mining_gold} can reduce the amount of simulated data needed by orders of magnitude. The Python library \toolfont{MadMiner}~\cite{Brehmer:2019xox} implements most of the machine learning--based algorithms discussed in Secs.~\ref{sec:surrogates} and \ref{sec:optimal_observables}. It wraps around \toolfont{MadGraph}, \toolfont{Pythia}, and \toolfont{Delphes} and thus automates the entire pipeline for a typical phenomenological study\footnote{The \toolfont{sbi} package~\cite{tejero-cantero2020sbi} implements many simulation-based inference methods, in particular for Bayesian inference, in a problem-agnostic way, but does not provide any interface to particle physics simulators yet.}. The approach is compatible with full simulation like \toolfont{Geant4} as the necessary information can just be passsed through the detector simulation similar to the weights used to assess uncertainty in the parton distribution functions. However, this still requires a modest investment in the experiments' simulation software. 

The use of the learned likelihood ratio for reweighting event samples has the potential for a significant reduction in simulation costs as the reweighting factor can often be learned on parton-level or particle-level data without running the full simulation or reconstruction on large samples of simulated data with varied parameter settings. The CARL technique is being explored within ATLAS and integrated into the ATLAS software framework for this purpose~\cite{leonora_vesterbacka_2020_4062049}.

Probabilistic programming also has the potential to address the computational resources needed for simulation at the high-luminosity LHC. Signs of new physics typically would hide in tails of background distributions, which are computationally expensive to populate with naive sampling approaches. HEP collaborations regularly use a form of importance sampling where the parton-level phase space is sliced (\eg slices in the transverse momentum of outgoing partons to fill the high-$p_T$ in the process $pp\to jj$). In this case, one merges several individual samples of simulated events weighted by the $N_s/\sigma_s$, where $N_s$ is the number simulated samples and $\sigma_s$ is total cross-section for that slice. However, this approach does not work for efficiently sampling regions of phase space that do not correspond to simple regions in the parton phase space. For example, if we want to populate the regions of phase space where standard QCD jets fake a boosted top tagger based on a deep neural network~\cite{Kasieczka:2019dbj} the fake rate is roughly $10^{-3}$ and much of the relevant fluctuations happen in the parton shower and are not reflected in the parton-level phase space. Event generators instrumented with probabilitic programming constructs offers the potential to efficiently  sample these complicated regions of phase space, which is being explored with a simplified parton shower known as \texttt{Ginkgo}~\cite{ginkgo}.

In the long term, we should not treat the simulation chain as a black box, but open them and begin to integrate automatic differentiation and probabilistic programming capabilities in them as that will enable more powerful and sample-efficient inference algorithms~\cite{Cranmer:2019eaq}.

%============================================================
\section{Summary}
\label{sec:summary}
%============================================================

Particle physicists have a suite of simulators at their disposal that can model essentially all aspects of particle collisions with impressive fidelity. These tools use Monte-Carlo methods to generate events, with the distribution of outputs depending on the parameters of the physics model. However, we cannot use these tools directly for inference because we cannot evaluate the probability the simulator to generate a specific observed event. Because the likelihood is intractable, we can not directly fit for the most likely parameter points or calculate exclusion limits from observed data. Historically, this challenge has been overcome by reducing the high-dimensional event data to one or two kinematic variables and to use histograms or analytic functions to model the distribution of these observables. This makes inference possible, but often degrades the sensitivity of the analysis.

Here we reviewed simulation-based (or likelihood-free) inference methods that allow us to infer parameters based on high-dimensional event data. These methods are closely connected to other important tasks in HEP and provide the ability to reweight events~\cite{cranmer_2016, Andreassen:2019nnm, leonora_vesterbacka_2020_4062049}, tune shower and detector-simulation parameters to data~\cite{Andreassen:2019nnm},  unfold distributions~\cite{Andreassen:2019cjw}, and anomaly detection~\cite{Andreassen:2020nkr}.
An important driver of these algorithms are the rapidly increasing capabilities of machine learning, which let us analyze high-dimensional data efficiently. In addition, extracting matrix-element information from the simulator and using it to augment training data can drastically reduce the number of simulations we need to run. We presented algorithms based on these two ideas in which a neural network is trained as a surrogate for the likelihood or the likelihood ratio function or defines optimal observables, which can then be used in a traditional histogram-based analysis.

In first phenomenological LHC studies, these algorithms have been applied to Higgs precision measurements in vector boson fusion~\cite{Brehmer:2018eca}, in $WH$ production~\cite{Brehmer:2019gmn}, and in $t\bar{t}H$ production~\cite{Brehmer:2019xox}, as well as $ZW$ measurements~\cite{Chen:2020mev} and the search for massive resonances decaying into dijets~\cite{Hollingsworth:2020kjg}. The new machine learning--based techniques consistently led to more sensitive analyses than traditional histogram-based approaches such as simplified template cross-section measurements~\cite{Brehmer:2019gmn}. With a range of diagnostic tools and ideas for uncertainty quantification available and software packages making the application of these methods easier, the application of these new simulation-based inference techniques to data collected at the LHC experiments seems imminent.

%------------------------------------------------------------
\section*{Acknowledgments}
%------------------------------------------------------------

We want to thank our collaborators Zubair Bhatti, Sally Dawson, Irina Espejo,  Joeri Hermans, Samuel Homiller, Felix Kling, Gilles Louppe, Siddharth Mishra-Sharma, Juan Pavez, Sinclert Perez, Tilman Plehn, and Markus Stoye. This work was supported by the U.\,S.~National Science Foundation (NSF) under the awards ACI-1450310, OAC-1836650, and OAC-1841471. We are grateful for the support of the Moore-Sloan data science environment at NYU.

%============================================================
% References
%============================================================

\bibliographystyle{tepml}
\bibliography{references}

%\blankpage
%\printindex[aindx]                 % to print author index
%\printindex                         % to print subject index

\end{document}